\documentclass[letterpaper]{JHEP3}

\usepackage{epsfig}
\epsfclipon

\def\Dsl{\hbox{/\kern-.6700em\it D}} 
\def\dsl{\hbox{/\kern-.5300em$\partial$}}

\def\eq{\begin{equation}}
\def\eeq{\end{equation}}
\def\eqa{\begin{eqnarray}}
\def\eeqa{\end{eqnarray}}

\def\bd{\begin{displaymath}}
\def\ed{\end{diplaymath}}

\def\Box{ {\,\lower 0.9pt\vbox{\hrule\hbox{\vrule height0.2cm \hskip 0.2cm \vrule height 0.2cm }\hrule}\,}}
\def\lsim{{\ \lower-1.2pt\vbox{\hbox{\rlap{$<$}\lower5pt\vbox{\hbox{$\sim$}}}}\ }}
\def\gsim{{\ \lower-1.2pt\vbox{\hbox{\rlap{$>$}\lower5pt\vbox{\hbox{$\sim$}}}}\ }}
          
\def\Dsl{\hbox{/\kern-.6700em\it D}} 
\def\dsl{\hbox{/\kern-.5300em$\partial$}}

\def\beginvector{\left( \begin{array}{c} }
\def\endvector{\end{array} \right)}
\def\endignore{}
\def\ignore#1\endignore{}

\def\S2{{\mathcal S}^2}

\def\S{\Sigma}

\newcommand{\be}{\begin{equation}}
\newcommand{\ee}{\end{equation}}
\newcommand{\bea}{\begin{eqnarray}}
\newcommand{\eea}{\end{eqnarray}}
\newcommand{\beqar}{\begin{eqnarray*}}
\newcommand{\eeqar}{\end{eqnarray*}}

\title{Cosmology of intersecting brane world 
models in Gauss-Bonnet gravity}

\author{
 Hyun Min Lee\footnote{minlee@th.physik.uni-bonn.de} \,and  
Gianmassimo Tasinato\footnote{tasinato@th.physik.uni-bonn.de}\\ {\it
 Physikalisches Institut der Universit\"at Bonn}\\  {\it Nussallee 12,
 D-53115 Bonn, Germany }}

\abstract{
We study the cosmological properties
of  a codimension two brane world that sits at the intersection
between two four branes, in the framework
of six dimensional Einstein-Gauss-Bonnet gravity. Due 
to contributions of the Gauss-Bonnet terms,
the junction
conditions require  the presence of  localized  energy density on the
codimension two defect.
 The induced metric on this surface assumes a FRW form,
with a scale factor associated to the position of the brane in the background;
 we can embed on
 the codimension two defect the preferred form of energy 
density.
We present the cosmological 
evolution equations for the three brane, showing
that, for the case of pure AdS$_6$  backgrounds, they acquire
the same form  of the ones for  the Randall-Sundrum II model.  When
the background is different from pure AdS$_6$, the  cosmological
behavior is potentially 
  modified in respect to the typical one of codimension
one brane worlds. 
We 
discuss, in a particular model embedded in an AdS$_6$ black hole,
the conditions one should satisfy  in order 
to obtain  standard cosmology at late epochs.}

\begin{document}

\section{Introduction}

\noindent {\it Status of cosmology on codimension two brane worlds}
\vskip 0.3 cm

\noindent
In the past few years,
the possibility that the observed world is localized on a brane, 
embedded in a higher dimensional  space, has received 
much attention, opening new possibilities for 
phenomenology~\cite{arkani,randall},
and furnishing interesting connections with string theory.

In particular,  cosmological aspects of brane-world models in codimension one
defects have been extensively analyzed,  due to the fact that, in this
case, it is possible to compute the backreaction of the brane  in the
bulk  geometry, using the so-called Israel-Lanczos junction
conditions. The most  studied models for cosmology, based on the
Randall-Sundrum II background,  are characterized by standard cosmological
 evolution  at late times,  while corrections to the standard Hubble equation,
due to modification of gravity,
occur at early  cosmological epochs.

\vskip 0.2 cm

The interesting properties
of brane world models  on codimension two defects
(for example, three branes on six dimensional space-times)
 have
  been recently re-considered, for the possibility to construct new 
examples of warped
 brane world scenarios in six dimensions \cite{branew6d},  or  to face 
  the cosmological constant problem from a new perspective \cite{cosmpos}
(see however 
\cite{cosmneg}).  Less attention has been dedicated to study cosmological
aspects of brane worlds  for codimension higher than one, since it is 
in general 
difficult to take into account in a consistent way
 the backreaction of the brane 
in this case. 
Relatively to this
 problem,
 it has been shown long ago
 \cite{hooft}
how to  embed consistently a brane, characterized
by  pure tension, on the tip of a conical singularity of the higher dimensional
geometry (see also \cite{gregoryp} for related examples).
The subsequent problem to embed 
more general 
energy density  on this brane  has been considered~\cite{giovannini},
with the quite surprising  result that, with the most natural
 ansatz  for the six dimensional metric, it is 
{\it not} possible to localize energy density that is different
from pure tension. A way out for this unpleasant situation
has been presented in \cite{bostock}, where the authors  observe that 
a Gauss-Bonnet term, added to  the initial six dimensional 
 action, 
 induces new terms on the brane junction conditions
that allow to  embed the preferred  matter on the conical defect.

\vskip 0.3 cm
\noindent {\it A new  approach}
\vskip 0.3 cm

\noindent
In this paper,  motivated by the ideas presented
in \cite{bostock} and \cite{HMLee},
we would like to present  an alternative approach to construct 
  cosmological models 
on  codimension two  defects, in the framework of  theories with 
Gauss-Bonnet terms. Our approach  is
 different from brane worlds 
 on the tip of a conical singularity:
in our scenario, a  codimension two defect (a three brane)
 is located
at the {\it intersection} between two codimension one, four branes at
right angle,  in a six
 dimensional space. Each of these four branes corresponds to a fixed 
hypersurface for a $Z_2$ symmetry, that identifies the space on the
two sides of the hypersurface.  The resulting geometry does
not present conical singularities.
We show that, in the framework of Gauss-Bonnet
gravity,  it is possible to consistently  work out the junction
conditions at the intersection, to
 localize  the preferred
energy density  on this codimension two defect,
and, in particular cases, 
 to obtain a (late time) standard cosmology, similarly
to    well-known  codimension one
examples in five dimensions.

The  main motivation  is to construct examples of
cosmological  models for our codimension
two brane worlds, in which also the background
geometry (bulk plus codimension one four branes that 
intersect) is completely specified.
 We have a
 complete control of the evolution equations  for the codimension
two brane world, and  we relate this  evolution to a
time dependence of  the system of codimension one branes.
This in general requires  a degree of fine tuning between energy
density  on the codimension two brane and the one on   the codimension
one branes. It
 would be interesting
to understand whether also other approaches  for constructing cosmological
models in codimension two defects  require similar fine tunings.

An important property of   our framework  consists on the fact
that it is not necessary
to consider  solutions
for Einstein-Gauss-Bonnet theory that present conical singularities. Indeed,
{\it any} bulk solution  is suitable to embed codimension
one  branes that intersect,
providing the codimension two brane-world we are interested in.
Moreover, our approach could suggest  a way to obtain  models for cosmology
based 
on the picture of intersecting brane worlds,
where the full backreaction of the branes is taken into account.
This kind of configurations receive large attention 
 in  String Theory,  for the possibility 
to obtain models with low energy  Standard Model-like spectra \cite{string2}.

\vskip 0.3 cm
\noindent {\it The models}
\vskip 0.3 cm

\noindent
We discuss  brane world models  defined
in vacua for  six dimensional Einstein-Gauss-Bonnet gravity with negative
cosmological constant~\footnote{
Brane-world models  embedded in  five dimensional vacua 
for Einstein-Gauss-Bonnet
 gravity have been  extensively studied \cite{onlylocal,pastGB}; less 
effort  has been devoted to construct models in
 six dimensions, with the exception of \cite{HMLee,Kakushadze}.}.
 We   embed in these space-times 
 two codimension one, four
branes that intersect at right angle, working out 
the  Israel-Lanczos   junction conditions. We show that, 
  in order to satisfy these  
conditions,  energy density   must be localized 
on    a codimension two defect located
at the  intersection between the four branes.
We focus on  
the case in which one of the four branes
{\it moves}
 through the static  AdS$_{6}$ background, in the  mirage cosmology framework
of \cite{kiritsis}, applied to $Z_2$ symmetric brane worlds
in \cite{kraus}. 
 The intersection between the branes moves
as well: we define our 3+1 dimensional brane world on this surface, 
and consequently we provide an example of codimension
two brane world that moves through the six dimensional
background.
 One finds that the energy density localized on the codimension two
brane  acquires
 different forms depending on how the energy momentum tensor
of the moving codimension one four brane is chosen. Looking on the other
way around, we can choose our preferred form for the energy density
localized 
on the intersection, and  consequently define the energy momentum tensor
on the four brane that moves.

In particular, we consider two models. The first one  is
a brane world system  embedded 
in a pure AdS$_6$ background: 
 working out in detail  the evolution equations
for the projected cosmology in the codimension two defect, we show that
the projected Hubble equation acquires {\it exactly} the same structure 
of the Hubble equation in the five dimensional Randall-Sundrum II model.
At late epochs the projected Friedmann equation  in the codimension 
two brane approaches the one of standard cosmology: this indicates that,
at large distances, gravity behaves as four dimensional, without
the necessity of compactifying the extra dimensions.  
We point out  that, as expected, 
a certain
degree of fine tuning is required between the energy 
density localized   on the three brane at the intersection, and the
 energy density    on the  moving   four brane. 
In the second model,  we use as background geometry an AdS black hole
in six dimensions. 
 We embed
also in this case two four branes that intersect
at right angle, and 
 we limit our attention to the cosmology 
on the codimension two intersection. The analysis of this more general 
 system  renders manifest
interesting differences between the evolution equations of our codimension 
two brane world, and the traditional  codimension one brane worlds. 
We find that,  in this 
case, the resulting cosmological properties are in general  non standard: 
the equation of continuity  for the  energy density is not satisfied, 
and the effective Planck mass varies with  time.
These non standard features are related to  the geometrical properties of the
background. We discuss the conditions to recover, at least at late epochs,
 a standard
cosmological  behavior.

\section{Formalism:  Branes in six dimensional Gauss-Bonnet
gravity}

Let us
  start presenting  the general  action for our six dimensional system

\begin{eqnarray}\label{mainaction}
S&=&M_{6}^4\int d^6x  \sqrt{-g}\bigg(\frac{1}{2}R-{\cal L}_{6}
+\frac{1}{2}\alpha {\cal L}_{GB}\bigg)\,,
\end{eqnarray}

\noindent
where
$M_{6}$, ${\cal L}_6$ and $\alpha$ are the 
six dimensional fundamental mass scale, 
the Lagrangian for bulk fields, and the Gauss-Bonnet coupling 
with mass dimension $-2$, respectively. The Gauss-Bonnet term
is defined in six dimensions as~\footnote{This 
term,  that in six dimensions does not 
correspond to a topological invariant,
  has been actually first discussed
 by Lovelock \cite{lovelok}. We thank V.~Rychkov for this terminological
remark.}
\begin{eqnarray}
{\cal L}_{GB}=R^2-4R_{MN}R^{MN}+R_{MNPQ}R^{MNPQ}.
\end{eqnarray}

\noindent
The  Einstein equations relative to this action are given by
\begin{eqnarray}
G_{MN}+2\alpha H_{MN}=T_{M N} \label{bulkeq}
\end{eqnarray}
with $T_{MN}$ the energy momentum tensor relative to the bulk 
Lagrangian ${\cal L}_6$, 
and 
\begin{eqnarray}
H_{MN}&=&RR_{MN}-2R_{MP}R^P\,_N-2R^{PQ}R_{MPNQ}+R_M\,^{PQR}R_{NPQR}
\nonumber \\
&-&\frac{1}{4}g_{MN}{\cal L}_{GB}.
\end{eqnarray}

\vskip 0.3 cm
\noindent
Considering  a  vacuum for this theory, we can embed two codimension
one branes (four branes)
$\Sigma_1$ and $\Sigma_2$ intersecting at right angle,
  corresponding  to fixed points
of two  $Z_2$ symmetries that identify
the space in each side of them. It is possible to 
localize fields in their world  volume, that are described  by 
the  action

\begin{eqnarray}
S_{br}=\int_{\Sigma_1} d^5 x \sqrt{-h_1}{\cal L}_1  
+\int_{\Sigma_2}d^5 x \sqrt{-h_2}{\cal L}_2
+\int_{\Sigma_3 \equiv \Sigma_1\cap\Sigma_2} d^4 x\sqrt{-h_3} {\cal L}_3.
\end{eqnarray}

\noindent
where $h_i$ denote the projected metrics
on the subspaces, and we include possible contributions localized
on a codimensions two defect at the intersection
between the branes.

The junction
conditions for each of the two codimension
one branes are dictated by the Israel junction conditions,
extended to the Gauss-Bonnet case in \cite{davis}. Let us present  the 
junction conditions on the brane $\Sigma_2$: the ones for the brane 
$\Sigma_1$ are analogous,  and both  sets of conditions
must be  satisfied simultaneously. 
They are
given by the equations

\begin{eqnarray}
2\langle K_{ab}-Kh_{ab}\rangle+2\alpha\langle 3J_{ab}-Jh_{ab}
+2P_{acdb}(h)K^{cd}\rangle 
=-\frac{1}{M_{6}^4}({\tilde S}_{ab}+S_{ab})\label{braneeq}
\end{eqnarray}
where  $\langle X\rangle \equiv 
(X(\Sigma_{2,+})+X(\Sigma_{2,-}))/2$,  while the  induced 
 four brane metric is 
$h_{MN}=g_{MN}-n_{M} n_{N}$, 
with a unit normal vector $n_{M}$ to this codimension one
 defect~\footnote{Capital 
letters denote 
 6d indices, $M,N=0,1,2,3,4,5$. Small letters denote 
 5d indices, $a,b=0,1,2,3,4$. Greek letters denote  4d
indices, $\mu,\nu=0,1,2,3$.}.
The  extrinsic curvature tensor evaluated on $\Sigma_2$  is given 
by $K_{ab}=h^M_{a} h^N_{b}\nabla_M n_N$ with $K=K^a\,_{a}$, and the
tensors $J_{ab}$ and $ P_{abcd}$  are given by

\begin{eqnarray}
J_{ab} &=& \frac{1}{3}(2KK_{ac}K^c\,_{b}
+K_{cd}K^{cd}K_{ab}-2K_{ac}K^{cd}K_{db}
-K^2 K_{ab})\,, \\ \nonumber\\
P_{abcd} &=& R_{abcd}+2R_{b[c}g_{d]a}-2R_{a[c}g_{d]b}+Rg_{a[c}g_{d]b}\,\,.
\label{divr0}
\end{eqnarray}

\noindent
The  energy momentum tensors relative
to matter localized on $\Sigma_2$ , appearing on the right hand side of
(\ref{braneeq}),  are given by

\begin{eqnarray}
{\tilde S}_{ab}&=&-\frac{2}{\sqrt{-h_2}}
\frac{\delta\left( \sqrt{-h_2}  {\cal L}_2\right)}{\delta h_2^{ab}}\,, \\
S_{ab}&=&-\delta(\Sigma_1)\delta_a^\mu\delta_b^\nu
\frac{2}{\sqrt{-h_2}}\frac{\delta \left(\sqrt{-h_3} 
{\cal L}_3\right)}{\delta h_3^{\mu\nu}}\equiv 
\delta(\Sigma_1) \sqrt{\frac{- h_{3}}{-h_2}}  \delta_a^\mu\delta_b^\nu 
S_{\mu\nu}\,\,.
\end{eqnarray}

The last quantity $S_{a b}$ denotes energy momentum
tensor that is  localized on the intersection $\Sigma_3$ 
between the branes. When considering the usual Einstein gravity,
one finds that, in this situation,
there are no  terms, in the geometrical
part of junction conditions, that 
are localized at the intersection. This means that 
 energy momentum tensor  $S_{ab}$ should  vanish.

This is no more true when we add the Gauss-Bonnet
part:
 contributions contained in the $P$-tensor, in 
the left hand side of (\ref{braneeq}),
  result to be localized just  at the intersection between the codimension
one branes, and require  energy momentum tensor at the intersection in order
to satisfy the junction conditions.
 We will show 
this fact 
by explicit examples  in the next sections.

\section{First     model: Intersecting branes in AdS$_{6}$}

In this section, we deal with the simplest example of
 bulk action in which the formalism presented in the previous section
can be applied: an AdS bulk, where ${\cal L}_6 \equiv
\Lambda_6$ corresponds
to negative cosmological constant, and  action given by

\begin{eqnarray}\label{mainaction2}
S&=&M_{6}^4\int d^4x dz_1 dz_2 \sqrt{-g}\bigg(\frac{1}{2}R-\Lambda_{6}
+\frac{1}{2}\alpha {\cal L}_{GB}\bigg) \nonumber \\
\end{eqnarray}

\noindent
Then, we can solve the  
 Einstein-Gauss-Bonnet
equations (\ref{bulkeq}), with $T_{MN}= -\Lambda_{6} g_{MN}$, obtaining
a warped metric with flat four dimensional subspace

\begin{eqnarray}\label{puremetr}
ds^2=A^2(z_1,z_2)(\eta_{\mu\nu}dx^\mu dx^\nu+dz_1^2+dz_2^2)\,,
\end{eqnarray}
with a  warp factor 
 given by \cite{HMLee}
\begin{equation}\label{solwarp}
A(z_1,z_2)=1/(k_1 z_1 +k_2 z_2 +1)\,,\end{equation}
while  the constants $k_1$ and $k_2$  satisfy the equation 
\begin{eqnarray}
k^2_1+k^2_2=\frac{1}{12\alpha}
\bigg[1\pm\sqrt{1+\frac{12\alpha\Lambda_{6}}{5}}\bigg] 
\,.
\label{bulk}
\end{eqnarray}
The two signs correspond to two branches of solutions: from now on, we 
consider the minus sign, since in this case taking $\alpha \to 0$ we obtain
the correct limit to the AdS solution of standard Einstein gravity 
\cite{kaloper}.

\vskip 0.2 cm

Let us embed in this background two codimension one branes, that intersect
with a 90 degree angle. In the next subsection, we review
 the conditions to have a static system, with the two static branes
  intersecting
at the origin. In subsection (3.2), instead, we
present a model in which one of the four branes moves through the bulk,
inducing cosmology both
 on its world-volume, and on the codimension two defect that lives
at the intersection  with  the other four brane.

\subsection{Static case}

Starting from metric (\ref{puremetr}), 
let us consider the situation in which the two four branes
are situated  respectively on  the hypersurfaces $z_1 =0$ and
$z_2 =0$~\footnote{This static system of branes at right angle
has been considered in
\cite{HMLee} (see also \cite{neupane}).}.
 Since they constitute fixed points of a $Z_2$ symmetry,
the bulk metric in this case becomes

\begin{eqnarray}
ds^2=A^2(|z_1|,|z_2|)(\eta_{\mu\nu}dx^\mu dx^\nu+dz_1^2+dz_2^2)
\end{eqnarray}
and the warp factor becomes
$A(|z_1|,|z_2|)=1/(k_1|z_1|+k_2|z_2|+1)$. The junction
conditions at the brane positions force
to choose the brane energy densities as pure tensions, 
that is:

\begin{eqnarray}
{\cal L}_1=-\Lambda_1\,,
 \ \ {\cal L}_2=-\Lambda_2\,, \ \ {\cal L}_3=-\lambda \,.
\end{eqnarray}

\noindent
The tensions must satisfy the relations

\begin{eqnarray}
k_1\bigg[1-12\alpha\bigg(\frac{1}{3}k^2_1+k^2_2\bigg)\bigg]
&=&\frac{\Lambda_1}{8M_{6}^4}\,, \label{t4a}\\
k_2\bigg[1-12\alpha\bigg(k^2_1+\frac{1}{3}k^2_2\bigg)\bigg]
&=&\frac{\Lambda_2}{8M_{6}^4}\,,\label{t4b}\\
\alpha k_1 k_2 &=& \frac{\lambda}{96M_{6}^4}\label{t3}\,.
\end{eqnarray}

\noindent
The interesting point is
 that, in this system,  the codimension two brane $\Sigma_3$
located at the intersection between the four branes
is characterized by its own tension
$\lambda$, proportional to the Gauss-Bonnet parameter $\alpha$.
This means that the presence, in the six dimensional action,
of the Gauss-Bonnet term implies the presence  of energy density 
localized on the three brane. We will develop this observation
in the context of moving branes in the next subsection.

\subsection{  Moving branes: Cosmology}\label{firstcosmo}

Starting from \cite{kiritsis}, it has been observed that
branes that move through  a static bulk induce cosmology on their
own world volume. The time evolving projected scale factor depends
on the position of the brane in the bulk, and the time dependent 
energy density localized
on the brane must satisfy  suitable junction conditions, that change with
changing  the brane position.

Let us apply this idea to our system, demanding that one of the two branes,
let us say $ \Sigma_2$, moves 
through  the static background constituted  by the AdS bulk, while  the other
brane $\Sigma_1$ is fixed at the position $z_1=0$. In this situation,
the  three brane that sits
on the intersection  moves as well,  and, as we will 
see, this generates an induced cosmology also at the intersection.

We start evaluating the junction conditions at the fixed points
of the $Z_2$ symmetries,  next we  continue discussing the features
of the projected cosmology  on the three and four branes.

\subsubsection{Evaluation of junction conditions}\label{jcfirst}

\noindent
We demand that the four brane $\Sigma_2$ moves through the static
AdS bulk in which the static $\Sigma_1$ brane is located. The starting
 bulk metric
is consequently  the following,

\begin{eqnarray}
ds^2=A^2(|z_1|,z_2)(\eta_{\mu\nu}dx^\mu dx^\nu+dz_1^2+dz_2^2)\,,
\end{eqnarray}

\smallskip
\noindent
and  the static brane $\Sigma_1$ located in $z_1=0$, the 
fixed point of a $Z_2$
symmetry,  satisfies the junction
condition (\ref{t4a}).
 Calling
$z_2(t)$ the position of the moving brane, 
 the induced metric on its world volume results to be

\begin{eqnarray}
ds^2_{\Sigma_2}&=&A^2(|z_1|,z_2(t))\bigg[-(1-{\dot z}_2^2)dt^2
+\delta_{ij}dx^i dx^j+dz^2_1\bigg] \nonumber \\
&\equiv & -n^2(|z_1|,t) dt^2+a^2(|z_1|,t)
 (\delta_{ij}dx^i dx^j+dz^2_1)\label{metric4b}
\end{eqnarray}
where the dot denotes the derivative with respect to $t$. 
Then, the  bulk space for $z_2 <z_2(t)$ is identified 
with the one for $z_2>z_2(t)$ by a $Z_2$ symmetry.

\smallskip

Let us continue evaluating the geometrical quantities
that appear in the left hand side of eq. (\ref{braneeq}), that
represent the junction conditions on the brane $\Sigma_2$.  The results
of our calculations  show that there are geometrical terms,
on this equation, that are localized at the intersection
between the four branes, and must be compensated by some
form of energy momentum tensor localized on this codimension two surface.

The velocity vector ($u^M$) and the normal vector ($n_M$) 
for the moving 4-brane are given by 
\begin{eqnarray}
u^M&=&\frac{1}{A\sqrt{1-{\dot z}_2^2}}(1,{\dot z}_2,{\vec 0})\,, \\
n_M&=&\frac{A}{\sqrt{1-{\dot z}_2^2}}(-{\dot z}_2,1,{\vec 0})\,.
\end{eqnarray}
Then, the spatial components of the extrinsic curvature at the 4-brane 
are given by
\begin{eqnarray}
K^{z_1}_{z_1}=K^i_i
&=&\frac{1}{2}n^{z_2}\partial_{z_2} h_{ij} \nonumber \\
&=&-\frac{k_2}{\sqrt{1-{\dot z}_2^2}}\equiv {\cal K}\,.
\end{eqnarray}
On the other hand, the $(00)$ component of the extrinsic curvature 
at the 4-brane is given by
\begin{eqnarray}
K^0_0=-K_{MN}\,u^M u^N=\frac{1}{A^2 {\dot z}_2}
\frac{d}{dt}\bigg(\frac{A}{\sqrt{1-{\dot z}_2^2}}\bigg)\,.
\end{eqnarray}
From the induced metric of the moving 4-brane $\Sigma_2$, eq.~(\ref{metric4b}),
we obtain the non zero components of
the Riemann tensor and of the Ricci tensor, respectively, 
as the following,
\begin{eqnarray}
R^{0i}\,_{0j}&=&(X-Y-{\bar Y})\delta^i_j, \hskip 1.5 cm
R^{0z_1}\,_{0z_1}=X-Y-{\bar X}+{\bar Y}, \nonumber \\
R^{ik}\,_{jl}&=&(Y-{\bar Y})(\delta^i_j\delta^k_l-\delta^i_l\delta^k_j), 
\hskip 0.6 cm
R^{iz_1}\,_{jz_1}=(Y-{\bar X}+{\bar Y})\delta^i_j,
\end{eqnarray} 
\begin{eqnarray}
R^0\,_0&=&4(X-Y)-{\bar X}-2{\bar Y}, \nonumber \\
R^i\,_j&=&(X+2Y-{\bar X}-2{\bar Y})\delta^i_j, \\
R^{z_1}\,_{z_1}&=&X+2Y-4{\bar X}+4{\bar Y} \nonumber
\end{eqnarray}
where we define
\begin{eqnarray}
X&=&\frac{1}{n^2}\bigg[\frac{\ddot a}{a}
-\frac{\dot a}{a}\bigg(\frac{\dot n}{n}-\frac{\dot a}{a}\bigg)\bigg]\,,
\hskip 1 cm 
Y = \frac{1}{n^2}\bigg(\frac{\dot a}{a}\bigg)^2\,, \\
{\bar X}&=&\frac{1}{a^2}\bigg(\frac{a^{\prime\prime}}{a}\bigg)\,,\hskip 3.1 cm 
{\bar Y} = \frac{1}{a^2}\bigg(\frac{a'}{a}\bigg)^2.
\end{eqnarray}
Then, the divergence-free part Riemann tensor, given
in  eq.~(\ref{divr0}), is given by
\begin{eqnarray}
P^{0i}\,_{0j}&=&(3Y-2{\bar X}+{\bar Y})\delta^i_j\,, \hskip 2.65 cm \ \ 
P^{0z_1}\,_{0z_1}=3(Y-{\bar Y})\,\,, \nonumber \\
P^{ik}\,_{jl}&=&(2X-Y-2{\bar X}+{\bar Y})
(\delta^i_j\delta^k_l-\delta^i_l\delta^k_j)\,, \hskip .1 cm\ \
P^{iz_1}\,_{jz_1}=(2X-Y-3{\bar Y})\delta^i_j\,\,. 
\end{eqnarray}

\vskip 0.4 cm

With these quantities, we  can evaluate the geometrical quantities
 of the  left hand side of the
junction conditions (\ref{braneeq}). We must now specify the form of 
the
energy  momentum tensors that appear in the right hand side of this equation.
We impose that the energy momentum tensors, both on the four brane
$\Sigma_2$ and on the three brane $\Sigma_3$,  acquire a perfect fluid form:

\begin{eqnarray} 
{\tilde S}^a\,_b
&=&{\rm diag}(-{\tilde \rho},{\tilde p},{\tilde p},{\tilde p},{\tilde p})\,,
\label{en4br}\\
\nonumber \\
S^\mu\,_\nu&=&{\rm diag}(-\rho,p,p,p)\,. 
\end{eqnarray}

\vskip 0.4 cm

Then, we can substitute all the quantities on the junction
conditions (\ref{braneeq}). The
 energy density and pressure on the {\it four}
 brane $\Sigma_2$, $\tilde\rho$ and $\tilde p$,
are given by

\begin{eqnarray}
{\tilde\rho}&=&-8M_{6}^4{\cal K}[1-4\alpha(3k^2_1+3k^2_2-2{\cal K}^2)],
\label{rho4}\\
{\tilde p}&=&2M_{6}^4(K^0_0+3{\cal K})[1-12\alpha(k^2_1+k^2_2)]
+64\alpha M_{6}^4 {\cal K}^3
-16\alpha M_{6}^4 \frac{z_2}{{\dot z}_2}\frac{d}{dt}{\cal K}^3,\label{p4}
\end{eqnarray}

\noindent
while the energy density and pressure on the {\it  three} brane $\Sigma_3$, 
$\rho$ and $p$, result to be

\begin{eqnarray}
\rho&=&-96\alpha k_1 M_{6}^4 {\cal K}|_{z_1=0}, \label{rho3}\\
p&=&-\rho+32\alpha k_1 M_{6}^4 (K^0_0-{\cal K})|_{z_1=0}\label{p3}.
\end{eqnarray}

\vskip 0.5 cm

\subsubsection{Cosmology at the intersection}

Since we obtained the connection between energy momentum
tensor  and geometrical quantities, we are ready to discuss
the cosmological properties of our background. We first consider  
 the 
induced cosmology on the codimension two, three brane $\Sigma_3$.

Starting from eq.~(\ref{metric4b}), a redefinition of the time variable allows
to write
 the induced metric on the 3-brane in a FRW form:
\begin{eqnarray}
ds^2_{3-brane}=-d\tau^2+R^2(\tau)\delta_{ij}dx^i dx^j\,,
\end{eqnarray}
\noindent
with
\begin{eqnarray}
R(\tau)\equiv A(z_1=0,z_2(\tau))=\frac{1}{k_2z_2(\tau)+1}\,,
\end{eqnarray}
where 
the proper time for the 3-brane is obtained from the 
 $t$ coordinate as 
a coordinate transformation, 
$$d\tau=A(z_1=0,z_2(t))\sqrt{1-{\dot z}_2^{2}}\, dt\,\,\, \rightarrow \,\,\,
  \sqrt{1-{\dot z}_2^{2}}=\frac{1}{\sqrt{1+A^{2}(\frac{d z_2}{d \tau})^{2}}}\,
.$$
The scale factor is related to the position of the codimension  two brane 
in the bulk.
Equations (\ref{rho3}) and
 (\ref{p3}) ensure that the standard continuity equation
is satisfied for energy momentum tensor on the three brane, that 
is,
\begin{eqnarray}
\frac{d\rho}{d\tau}+3H(\rho+p)=0\,.
\end{eqnarray}

Equation (\ref{rho3}) furnishes us the Friedmann equation on the brane,
that acquires the remarkably simple form 

\begin{eqnarray}
H^2=k^2_2\bigg[\bigg(\frac{\rho}{\lambda}\bigg)^2-1\bigg]\label{hubble}\,.
\end{eqnarray}

\noindent
where the tension $\lambda$ is defined in the eq.~(\ref{t3}).
It is easy to show that (\ref{hubble}) can be re-casted on the same form 
of  Friedman equation that one finds
in the Randall-Sundrum model. Indeed,  decomposing
\begin{equation}
\rho \equiv \lambda+\rho_{3}\label{decompo}\,,\hskip 1 cm p \equiv 
-\lambda+p_{3}\,,\end{equation}
and defining
\begin{equation}
M_4^{2} = \frac{\lambda}{6 k_2^2} = \alpha \frac{16  k_1 M_6^4}{ k_2}\,,
\label{m4}
\end{equation}
one can rewrite eq.~(\ref{hubble}) as
\begin{equation}\label{hubble2}
H^2=\frac{1}{3M_4^2} \rho_3 +\frac{1}{6M_4^2} \frac{\rho_3^2}{\lambda}\,.
\end{equation}

\smallskip

Therefore, we see that in this model
 we obtain standard cosmology at late 
times, when $\lambda \gg \rho_3$,  while we have corrections to the standard
form of the Hubble parameter at early times.
The fact that at late epochs the second term  in the right hand side of
 (\ref{hubble2}) becomes
negligible, and we obtain the standard form for the Friedmann equation,
indicates that gravity on the brane behaves as four dimensional at large
distances. Interestingly,  notice that in this model the four dimensional,
induced Planck mass, given by (\ref{m4}), results proportional to 
the Gauss-Bonnet parameter $\alpha$, like in the approach of \cite{bostock}.

As anticipated, the
 form
of the Friedmann equation (\ref{hubble}) is the same as the codimension
one, Randall-Sundrum II brane world. This observation will be true 
only 
when the background is purely AdS: for a  more general bulk, as we will
discuss in Section (\ref{secsecond}),
 the evolution equations are  different
in respect to the ones that typically  arise in  codimension one brane worlds.

\subsubsection{Cosmology on the four brane}

Following the same procedure of the previous case, 
we can write the cosmological evolution equation
for the moving four brane $\Sigma_2$, with induced
metric given in eq. (\ref{metric4b}). Equations
(\ref{rho4}) and (\ref{p4}) ensure that a continuity equation
for energy momentum tensor on the four brane (given in eq. (\ref{en4br}))
is satisfied:
\begin{eqnarray}
{\dot {\tilde \rho}}+4\frac{\dot a}{a}({\tilde \rho}+{\tilde p})=0\,,
\end{eqnarray}
where we use here the  six dimensional time $t$.

The form of the Friedmann equation on the brane is more
 complex than the three brane case, and contains
various contributions coming from the Gauss-Bonnet terms.
It is convenient to define a (position
dependent) proper time $\tilde{\tau}$ via the formula
\begin{equation}
d \tilde{\tau}=A(z_1,z_2(t)) \sqrt{1-\dot{z}_{2}^{2}} d t\,,
\end{equation}
where notice that for the limiting case $z_1=0$ this expression
coincides with the one we used for  the three brane proper 
time.  Defining the  four brane scale factor
as 
$$ R(\tilde{\tau})=A(z_1, z_2( \tilde{\tau}))\,,$$
we obtain the following form  for the four brane Friedmann equation:

\begin{equation}\label{fri4bran}
H^2= \left( \frac{1}{R} \frac{d R}{d \tilde{\tau}} \right)^{2}
 = c_{+}+c_{-} -\frac{1}{12 \alpha}\sqrt{1+\frac{12 \alpha \Lambda_{6}}{5}}
- k_2^2\,\,.
\end{equation}

\vskip 0.3 cm
\noindent
The quantities $c_{+}$ and $c_{-}$ are given by the following expressions

\begin{equation}
c_{\pm}=\frac{1}{24 \alpha} \left[ (1+ \frac{12 \alpha 
\Lambda_{6}}{5})^{\frac{3}{2}} +\frac{27 \alpha}{16 M_{6}^{8}} 
\tilde{\rho}^{2} \pm \frac{9}{2} \sqrt{\frac{\alpha}{6}}\frac{\tilde{\rho}}{
 M_{6}^4}\sqrt{(1+\frac{12 \alpha \Lambda_{6}}{5})^{\frac{3}{2}}+
 \frac{27\alpha}{32 M_6^8}\tilde{\rho}^{2}} \right]^{\frac{1}{3}}\,\,.
\end{equation}

\vskip 0.2 cm

It is not easy to extract interesting physical information
from this Friedmann equation. In any case,
decomposing the energy density and the pressure as
\begin{equation}
\tilde{\rho} 
\equiv \Lambda_2 +\rho_{4}\label{decompo2}\,,\hskip 1 cm \tilde{p}
 \equiv -\Lambda_2+p_{4}\,,\end{equation}

\noindent
and inserting  these  expressions in the right
hand side of
eq. (\ref{fri4bran}), one finds that, for $\Lambda_2 \gg \rho_4$, 
the dominant contribution    is a {\it linear} term in the
energy density $\rho_4$.

\subsubsection{Fine tuning issues}
The  motion    of the four brane through
the bulk  is necessarily
related to the one of  the three brane,
 since the latter sits on the former.  
This means that,
in  the limit $z_1 \to 0$, 
 we should impose  that the evolution for the
 four brane  scale factor, ruled by the corresponding 
Friedmann equation,
 results 
the same as  the one for the three brane scale factor 
(that
 is, the intersection must ``follow'' the moving four brane).

Calling (${\tilde \rho}_0,{\tilde p}_0$) energy density 
and pressure 
for the four brane at $z_1=0$, straightforward calculations
using the junction conditions 
show  that we must impose the
following relations between  energy densities and pressures 
 between the three and the four brane:

\begin{eqnarray}\label{ftcond1}
{\tilde \rho}_0&=&\frac{\rho}{\lambda}\bigg[\Lambda_2-64\alpha k^3_2 M_{6}^4
\bigg(1-\frac{\rho^2}{\lambda^2}\bigg)\bigg],\\
{\tilde p}_0&=&\frac{1}{4}{\tilde \rho}_0\bigg(-1+\frac{3p}{\rho}\bigg)
+96\alpha k^3_2 M_{6}^4 \bigg(\frac{\rho}{\lambda}\bigg)^3
\bigg(1+\frac{p}{\rho}\bigg)\,\,.\label{ftcond2}
\end{eqnarray}

These equations represent fine tuning relations that must be satisfied,
in our system, in order to  obtain consistent cosmology. 
It is interesting to see that, at late epochs of cosmological
evolution, they acquire a simple, linear
 form. Indeed, decomposing the energy densities and pressures
as in eqs. (\ref{decompo}) and (\ref{decompo2}),
the fine-tuning equations (\ref{ftcond1})-(\ref{ftcond2}) 
become, at linear order,

\begin{eqnarray}
\rho_{4}&\simeq& (\Lambda_2+128\alpha k^3_2 M_{6}^4)
\frac{\rho_3}{\lambda},
\\
p_{4}&\simeq& -{\rho}_{4}
+\frac{3}{4}\bigg(\Lambda_2+128\alpha k^3_2 M_{6}^4\bigg)
\frac{1}{\lambda}(\rho_3+p_3)\,\,.
\end{eqnarray}

\vskip 0.2 cm

The ratios between pressure and energy densities, that define
the equations of state, are related, in this late epoch  limit,
via the following  expression

\begin{eqnarray}
\frac{p_{4}}{\rho_{4}}
\simeq\frac{3}{4}\bigg(\frac{p_3}{\rho_3}-\frac{1}{3}\bigg)\,\,.
\end{eqnarray}
This is the expression one expects to obtain
when the Friedmann equations for the three and the four brane are dominated
by the term linear  in the respective energy density.

\vskip 0.4 cm
\section{Second  model: Intersecting Branes in AdS Black Hole}\label{secsecond}

In Einstein-Gauss-Bonnet theories with negative cosmological
constant, the  most general solution for the equations
of motions relative to the system of eq. (\ref{mainaction2}),
with a spherical metric ansatz, 
is given in \cite{bhsol}. This is given by  
\begin{equation}\label{ansBH}
d s_{6}^{2}=-h(r) d t^{2}+\frac{d r^{2}}{h(r)}+r^{2} d \Omega_{4}^{2} \,,
\end{equation}
where the metric coefficient is given by
\begin{equation}\label{solBH}
h(r)=q+\frac{r^{2}}{12 \alpha} \left(1\pm\sqrt{1+\frac{12 \alpha}{5}\Lambda_6
+\frac{24\alpha}{r^{5}}\mu} \right)\,,
\end{equation}
where $q$ represents the curvature of the four dimensional
subspace labeled by $d\Omega_4$, and $\mu$ is a mass parameter~\footnote{
This metric ansatz is  also suitable  to describe solutions
for different systems,
 containing for example antisymmetric forms \cite{cvetic}.}.
From now on, we will consider the solution branch with the minus
sign in (\ref{solBH}): in this case, the limit $\alpha \to 0$ corresponds
to an AdS-Schwarzschild geometry of normal Einstein gravity.
In the limit $\mu \to 0$, and $q=0$, one finds the same background discussed
in the previous section.

We would like to use this
general  background to embed a pair of intersecting
codimension one branes, and  study the cosmology at the intersection 
along the same lines of the model discussed in the previous section.
In order  to have  a more natural embedding of codimension one branes
intersecting at right angle, and to allow a direct comparison 
  with the  previously discussed 
model,  it is convenient to
  change coordinate system in the following way~\footnote{We will 
use again the coordinate system of (\ref{ansBH}) at the end of the
next subsection, since it allows to present the results in a particularly 
transparent form.}.
Defining $\frac{d r}{d \chi}=r \sqrt{h(r)}$, the metric (\ref{ansBH}),
for the case $q=0$,
can be re-casted in the form
\begin{equation}\label{secformBH}
d s_{6}^{2}=-h(\chi) d t^{2}+r^{2}(\chi) \left( \delta_{ij}
d x^{i} dx^{j}+d\chi^{2}+d\eta^{2}
\right) \,.
\end{equation}
Now,  let us re-define the extra coordinates as  
$$
\chi= \left(1+k_1 z_1+ k_2 z_2 \right)\,, \hskip 1 cm
\eta=\left(1+k_2 z_1- k_1 z_2 \right)\,,
$$
where $k_1$ and $k_2$ are two constants
satisfying the relation (\ref{bulk}). We
can rewrite the metric (\ref{secformBH}) as
\begin{equation}\label{thirdformBH}
d s_{6}^{2}=-B^{2}(z_1,z_2)d t^{2}+ A^{2}(z_1,z_2) \left( \delta_{ij}
d x^{i} dx^{j}+d z_{1}^{2}+d z_{2}^{2} \right)\,.
\end{equation}
The metric coefficients $A$ and $B$ are  defined in terms
of $h$ and $r$ as:
$$
B^2 \equiv h(z_1,z_2)\hskip 0.5 cm ,\hskip 0.5 cm  A^2 \equiv (k_1^2+k_2^2) 
r^2(z_1,z_2)\,.
$$
 For the limiting case $\mu=0$, it is easy to see
that this form of 
the metric reproduces (\ref{solwarp}), that is,
$$
A = B = \frac{1}{1+k_1 z_1+ k_2 z_2 }\,.
$$

\vskip 0.5 cm

\subsection{Cosmology at the intersection}

Starting with the metric (\ref{thirdformBH}), it is straightforward 
to embed a pair of codimension one branes. Similarly to what we have
done 
 in Section (\ref{firstcosmo}), we ask that one four brane $\Sigma_1$ 
 sits on the line $z_1=0$, the fixed point of an orbifold 
symmetry. The starting bulk 
 metric is consequently
\begin{equation}
d s_{6}^{2}=-B^2(|z_1|,z_2)d t^2+ A^{2}(|z_1|,z_2) \left( \delta_{ij}
d x^{i} dx^{j}+d z_{1}^{2}+d z_{2}^{2} \right)\,.
\end{equation}

\vskip 0.1 cm

At this point, 
we introduce
 a 
second, dynamical four brane $\Sigma_2$, that
intersects with a right angle  the first one, 
and moves through the bulk.  The induced metric on 
$\Sigma_2$ results to be
\begin{eqnarray}
ds^2_{\Sigma_2}
 = -\left[ B^2\Big(|z_1|,z_2(t)\Big)
- A^2\Big(|z_1|,z_2(t) \Big) {\dot z}_2^2 \right] dt^2
+A^2\Big(|z_1|,z_2(t)\Big)\left(\delta_{ij}dx^i dx^j+dz^2_1 \right)
\end{eqnarray}
where the dot denotes the derivative with respect to $t$. 
Then, the  bulk space for
 $z_2 <z_2(t)$ is identified 
with the one for $z_2>z_2(t)$ by a $Z_2$ symmetry.

The
evaluation of the junction conditions for the four branes 
and for the three brane at the intersection corresponds 
to a straightforward generalization of calculations developed
in Section (\ref{jcfirst}): in this case, one generally
 finds that the energy
density on the  four brane should  be anisotropic, since it depends
on the coordinate $z_1$. 
 In the present paper, we limit our analysis
to  the results for   the evolution equations
of the three brane at the intersection.

\vskip 0.1 cm

The induced metric on the three brane takes 
of the FRW form
\begin{equation}
d s_{3+1}^{2}=-d \tau^{2} +R^{2}(\tau) \left(\delta_{ij}dx^i dx^j\right)\,,
\end{equation}
where the scale factor again depends on the position of the 
three brane in the background, and is given by
\begin{equation}  R(\tau) \equiv (k_1^2+k_2^2)^{\frac{1}{2}} \, r(\tau)= 
 A\left(0,z_2\Big(t(\tau)\Big)\right)\,.\label{scfacsecmod2}\end{equation}
The proper time is connected to the original time $t$ via the definition
$$
d \tau = \sqrt{B(0,z_2)^{2}-  A(0,z_2)^{2}\, \dot{z}_{2}^{2}}\, d t\,.
$$

\smallskip

\noindent
The equations that rule the cosmological evolution
for the system are the equation of conservation of energy, and the
Friedmann equation. 
We consider a 
three brane  energy momentum tensor of the perfect fluid form
$$S^\mu\,_\nu = {\rm diag}(-\rho,p,p,p)\,. 
$$

\noindent
The standard continuity equation is {\it not} satisfied in the 
present case:
\begin{equation}\label{noncons}
\frac{d \rho}{d \tau}+3 \frac{1}{R(\tau)}\frac{d R(\tau)}{d \tau}
 (\rho+p) =  \rho \frac{d}{d \tau}
\ln  \frac{\sqrt{h(r(\tau))}}{ (k_1^2+k_2^2)^{\frac{1}{2}} r(\tau)}\,,
\end{equation}
where  we express this, as well as the following results, in terms of
the metric function $h(r)$ as defined in eq. (\ref{solBH}).
It is clear that in this frame  a flow of energy leaves the intersection
where the three brane is located.
Notice that in the case $\mu = 0$, 
\begin{equation}\label{intlim}
\sqrt{h(r)} \propto r\,,
\end{equation}
the right hand side of  (\ref{noncons}) vanishes, and we recover the
result of pure AdS. Interestingly,
starting from the expression for $h(r)$,
 one can render the right hand side of (\ref{noncons})
small enough 
imposing bounds on the
size of $\mu$, and/or requiring that the size of the scale factor
is large enough.

\vskip 0.2 cm

Similar features  occur  discussing the 
 Friedmann equation, that in this model results to be 
(remember the linear relation between $R$ and $r$, given in eq. 
(\ref{scfacsecmod2}))
\begin{equation}\label{frsecmod}
H^{2}\equiv \left(\frac{1}{R}{\frac{d R}{d \tau}} \right)^{2}
= k_2^{2} \left[\frac{
R^{2}}{h(r)} \left(\frac{\rho}{\lambda}\right)^{2}
-\frac{h(r)}{ R^{2}} \right]\,,
\end{equation}
where we use eq. (\ref{t3})  to define the constant
$\lambda$.
 A static configuration is obtained choosing, for
example,  $\rho=\lambda$, and 
tuning the scale factor (that is, the position
of the brane in the bulk) in such a way that the right hand side of 
(\ref{frsecmod}) vanishes.
Next,   cosmological
evolution is induced 
perturbing the static configuration with $\rho =\lambda+\rho_3$,
and writing
\begin{equation}\label{freqmod}
H^2 = k_2^{2}  \frac{R^{2}}{h(r)}\frac{1}{\lambda}
\left[2   \rho_3+ \frac{\rho_{3}^2}{\lambda} + \lambda-
\lambda \left(\frac{h(r)}{ R^{2}}
\right)^{2}
 \right]\,.
\end{equation}

\noindent
In this way, one recovers a linear term in the energy density
plus corrections. Notice that, in this case,
the
 effective Planck mass in four dimensions depends on the value of the
scale factor (unless $\mu=0$),  being given by
\be\label{secplmass}
M_{Pl}^{2}=\frac{\lambda}{6 k_{2}^{2} }
\frac{h\left(r(\tau)\right)}{R^{2}(\tau)}\,.
\ee

\noindent
In this case the same arguments presented in the case of
the equation of continuity are still valid. In particular, one  can render
the variation rate of the Planck mass, and the size of the corrections
in the Friedmann equation, small enough imposing bounds on the size of
$\mu$. 
Let us consider, for example, 
 an  expansion of the Friedmann equation  for small energy densities
and small $\mu$  parameter. 
With $p=-\lambda+p_3$ and $p_3=\omega_3\rho_3$, 
we get 

\begin{equation}
H^2\simeq \frac{2k^2_2}{\lambda}{\bar\rho}_3
+\frac{2(1+6\omega_3)k^2_2}{5(-2+3\omega_3)}\frac{\tilde\mu}{r^5}
+{\cal O}\bigg(\frac{{\bar\rho}^2_3}{\lambda}\bigg)
+{\cal O}\bigg(\frac{\tilde\mu}{r^5}{\bar\rho}_3\bigg)
+{\cal O}\bigg(\frac{{\tilde\mu}^2}{r^{10}}\bigg)
\end{equation}
where we define
\begin{eqnarray}
{\bar\rho}_3&\equiv& \rho_0 R^{-3(1+\omega_3)}, \ \ {\rm with}\ 
{\rm constant}\ \rho_0, \\
{\tilde\mu}&\equiv&\frac{5}{2}
\frac{\mu}{(k^2_1+k^2_2)\sqrt{1+12\alpha\Lambda_6/5}}. 
\end{eqnarray}
Therefore, we find that the correction due to the 
nonzero $\mu$ reflects a form of  dark radiation on the moving 4-brane.

\smallskip 

Let us end commenting on  the exact forms for the evolution equations,
(\ref{noncons}) and (\ref{frsecmod}),  that we obtain in this model.
These equations are written
in such a   way that they are valid for 
{\it any} bulk metric  of the form (\ref{ansBH}), and 
point out interesting differences between the results of our
 approach, and typical results found in  codimension one brane worlds.
For example, the factor of the $\rho^2$ term in eq. (\ref{frsecmod}),
related to the Planck mass in four dimensions,  depends
in general 
not only on the 
volume of the extra dimensions, like in the codimension one case,
but also on the position of the brane in the bulk. A similar observation
 regards the non vanishing  
flow of energy density from the brane to the background: 
in this frame, we obtain the standard continuity equation 
 only in the case of  pure AdS$_6$ background.
However, notice that  it is   straightforward 
 to Weyl rescale the metric
to another frame, where  we can  recover a 
standard continuity equation.

\section{Conclusions}

We studied the cosmological properties
of  codimension two three brane worlds that sit at the intersection
between two codimension one four branes, in the framework
of six dimensional Einstein-Gauss-Bonnet gravity.
The full backreaction of the defects in the background
is completely taken into account. We  showed that the junction
conditions force the presence of  localized  energy density on the
codimension two defect: the induced metric on this surface assumes a FRW form,
with a scale factor associated to the position of the brane in the background.
We presented the cosmological 
evolution equations for the three brane, showing
that, for the case of pure AdS$_6$  backgrounds, they acquire
the same form  that one finds in the Randall-Sundrum II model.  In particular,
we can embed on the codimension two defect the preferred form of energy 
density.
The fact
that we obtain 
standard cosmology at late times indicates that gravity on the brane 
behaves as four dimensional at large distances, with a Planck mass
that is proportional to the Gauss-Bonnet parameter $\alpha$.
 These properties are
obtained at the
price of fine tuning relations between the energy density localized on the
three brane, and energy  that sits on one of the four brane. When
the background is different from pure AdS$_6$, the  cosmological
behavior is potentially 
  modified in respect to the typical one of codimension
one brane worlds, due to the presence
of additional contributions to the evolution
equations. These contributions depend on the geometrical properties
of the background, and in general describe a flow of energy 
from the brane to the background. We 
discussed, in a particular model,
the conditions one must satisfy 
to obtain  standard cosmology at late epochs.

\smallskip

Although we considered examples in which the 
background  geometry is controlled just by negative 
cosmological constant, our approach is completely general,
and can be used to embed codimension two brane 
worlds in vacua for Einstein-Gauss-Bonnet gravity
coupled to other fields. This suggests the possibility 
that a coupling between brane matter and background fields, like for 
instance bulk scalars, 
 modify the  evolution equations, for example
compensating the energy flow from brane to bulk,  
and could also
 indicate a possible way out  for the fine tuning problems we
discussed.

Moreover, in principle,   
a suitable generalization of our approach
to systems that contain higher order terms in curvature invariants,
can provide a framework to study cosmology for brane worlds in codimension
higher than two.

\acknowledgments{
We are pleased to thank S.~F\"{o}rste, R.~Gregory,
N.~Kaloper, H.~P.~Nilles, 
A.~Papazoglou, and I.~Zavala
for discussions and comments on the manuscript. 
This work is supported by the
European Community's Human Potential Programme under contracts
HPRN-CT-2000-00131 Quantum Spacetime, HPRN-CT-2000-00148 Physics Across the
Present Energy Frontier and HPRN-CT-2000-00152 Supersymmetry and the Early
Universe. HML was supported by priority grant 1096 of the Deutsche
Forschungsgemeinschaft.}


\end{document}